\documentclass[twocolumn,english,prl,showpacs]{revtex4-1}
\usepackage[latin9]{inputenc}
\usepackage{amstext,bm}
\usepackage{graphicx}
\usepackage{xcolor}
\usepackage{amsmath}
\usepackage[titletoc]{appendix}
\usepackage{hyperref}
\usepackage{float}

\begin{document}

\title{ Topological states of non-Dirac electrons on triangular lattice}

\author{Qi-Feng Liang$^{1,2}$, Rui Yu$^{2}$, Jian Zhou$^{3}$ and Xiao Hu$^{2}$}
\email{Hu.Xiao@nims.go.jp}

\address{$^{1}$ Department of Physics, Shaoxing University, Shaoxing 312000,
China}

\address{$^{2}$ International Center for Materials Nanoarchitectornics (WPI-MANA)
National Institute for Materials Science, Tsukuba 305-0044, Japan}

\address{$^{3}$ Department of Materials Science, Nanjing University, Nanjing
210093, China}

\begin{abstract}

We demonstrate the possibility of topological states for non-Dirac electrons.
Specifically it is shown that, because of the  $C_{\rm 3}$ crystal symmetry and time reversal symmetry, $p_x$ and $p_y$ orbits accommodated on
triangular lattice exhibit a quadratic band touching at $\Gamma$ point at the Fermi level.
When the atomic spin-orbit coupling (SOC) is taken into account, a gap is opened resulting in a quantum spin Hall effect state.
As revealed explicitly by a $k\cdot p$ model, the topology is associated with
a meron structure in the pseudo spin texture with vorticity two, a mechanism different from honeycomb lattice and the band inversion.
One possible realization of this scheme is the 1/3 coverage by Bi atom
adapted on the Si{[}111{]} surface. First-principle calculations are carried out, and a global gap of $\sim 0.15$eV is observed.
With the Si substrate taking part in realizing the nontrivial topology, the present template is expected to
make the integration of topological states into existing electronics and photonics technologies promising.
\end{abstract}

\date{\today}


\maketitle

\noindent\textit{Introduction ---}
Topology in condensed matters has been attracting significant interest in recent years
\cite{Klitzing_1980,
DasSarma_Nayak_2008,
Kitaev_2009_2,
Hasan_Kane_RMP_2010,
Qi_zhang_RMP_2011}. Because of the
bulk-edge correspondence, topological systems exhibit robust surface states. The surface current of a topological
insulator is dissipationless, which may resolve the issue of Joule heating in nanoscale
electronics, and the Majorana quasi-particle excitations of topological superconductor may be exploited for
decoherence-free quantum computation.

By now topological states without strong external magnetic field
have been proposed theoretically and some are realized experimentally in various systems
\cite{
Haldane_1988,
Kane_Mele_Graphene_PRL_2005,
Kane_Z2_2005,
Bernevig_HgTe_QW_science_2006,
Molenkamp_HgTe_science_2007,
Zhang_bi2se3_2009,
TI_exp_Yazdani_2009,
TI_exp_XueQK_2009,
TI_exp_Hasan_2009,
TI_exp_SZX_2010,
Hasan_Kane_RMP_2010,
Qi_zhang_RMP_2011,
DuRuiRui_InAs_GaSb_2011,
TI_exp_Yazdani_2_2011}.
The topology of electronic bands is characterized by non-zero Chern
number for time-reversal symmetry broken systems\cite{TKNN_1982}, and spin Chern
number (elaborated as $Z_2$ index) in time reversal symmetric systems\cite{Kane_Mele_Graphene_PRL_2005,Kane_Z2_2005,FuLiang_3DTI_2007}.
It is illuminating to understand Chern numbers in various systems
in terms of topological texture of pseudo spin in momentum space.
On honeycomb lattice, the pseudo spin referring to the two sublattices
exhibits two merons at the Dirac cones at $K$ and $K'$ points,  each
contributing a topological charge 1/2. As seen in the Haldane model and
Kane-Mele model \cite{Haldane_1988,Kane_Mele_Graphene_PRL_2005,Kane_Z2_2005},
topological states such as quantum anomalous Hall effect (QAHE) and quantum spin Hall effect (QSHE)
can be achieved by tuning the signs of mass terms at $K$ and $K'$ points \cite{LWH_2013}.
Bernevig, Hughes and Zhang revealed that in the HgTe/CdTe quantum well a band inversion takes place
at $\Gamma$ point and generates a QSHE state, where the Dirac property sets in
due to the opposite parities of the two relevant electronic bands\citep{Bernevig_HgTe_QW_science_2006}.
In this case, the pseudo spin referring to the two orbits develops
a skyrmion in momentum space  with topological charge of unity.
In both schemes, the interactions between two pseudo spin states are
linear with momentum. Partially because of this historical reason, topological states
are sometimes understood as a unique property of Dirac electrons.

In the present work, we address the topology of non-Dirac electrons taking into account the effect of crystal symmetry and
spin-orbit coupling (SOC). It is revealed that, featured by quadratic, non-Dirac dispersions,
the relevant pseudo spin texture is a single meron with vorticity two carrying a topological charge of unity.
As a specific case, we consider the Si{[}111{]} surface with 1/3 coverage by Bi atoms.
In this system, Bi atoms are adapted regularly on the top layer of Si
atoms and form a triangle lattice with large unit cell of $\sqrt{3}\times\sqrt{3}$, characterized by the $C_{\rm 3v}$ symmetry.
The $p_z$ orbits of Bi atoms and those of Si atoms beneath them form a strong
bonding state, and become inactive. The $p_x$ and $p_y$ orbits of
Bi atoms and the $p_z$ orbits of the other two uncovered Si atoms remain
active. At $\Gamma$ point, the $p_x$ and $p_y$ orbits
are close to the Fermi level, which with SOC contribute to form a QSHE insulating state in a
unique way featured by the $C_{\rm 3v}$ crystal symmetry as well as time reversal symmetry.
As a second example, regularly adapted Bi, Au and H atoms on the SiC{[}111{]} surface are investigated.
While the crystal symmetry is lowered to $C_3$, similar physics is available as supported by time reversal symmetry.

It is intriguing to notice that topological gaps in the present systems are large, since the
strong atomic SOC\cite{XuYong_ZhangSC_BiH_PRL_2013,LiuFeng_BiH_Si_2014,Yao_BiX_SbX_2Dhoneycomb,CJW_2014} of Bi atoms
contributes totally in gap opening. The Si substrate is crucial in providing the
crystal symmetry and selecting the orbits of adapted Bi atoms desired for topological states. Therefore, the present
template makes it very promising to integrate topological states into existing electronics and photonics devices.

\begin{figure*}[]\protect
\centering
\includegraphics[width=0.8\textwidth]{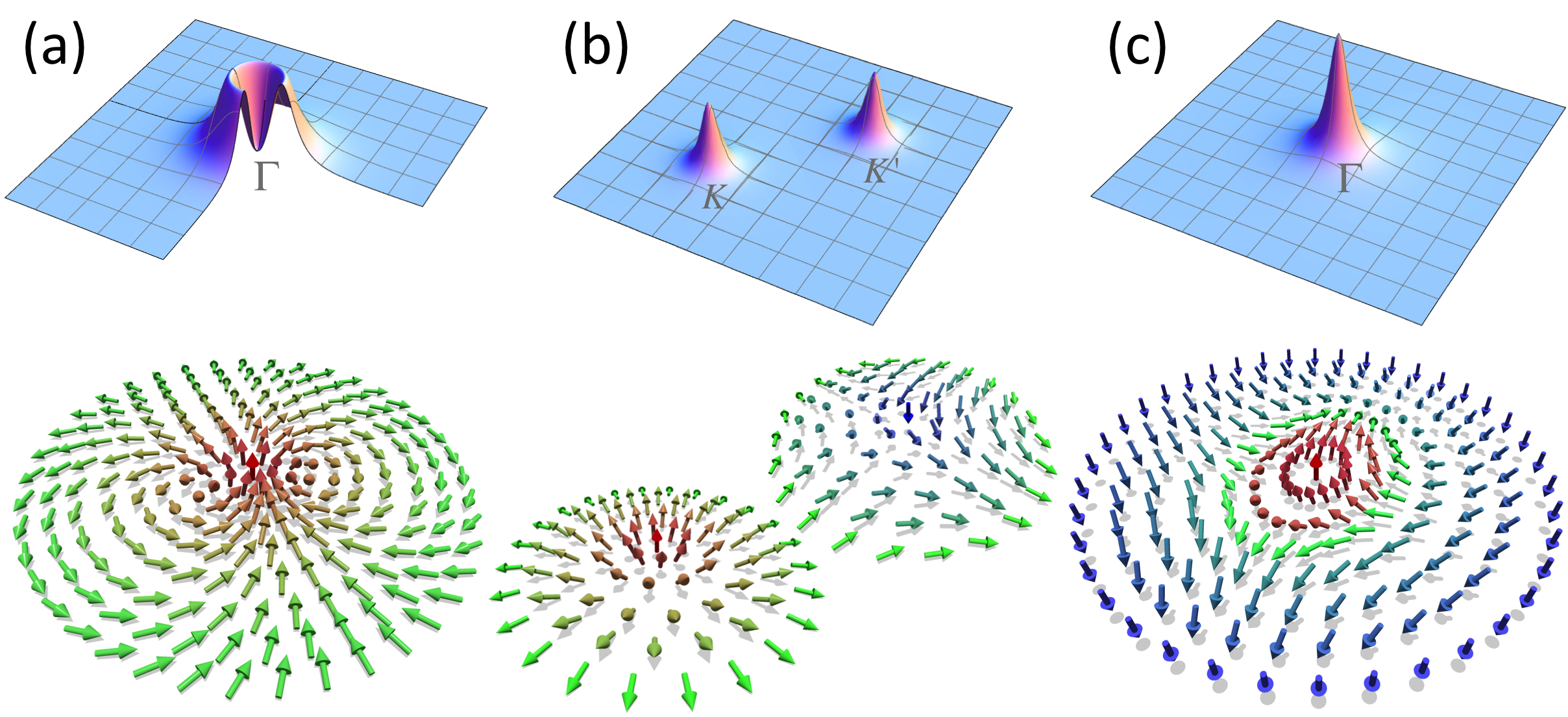}
\caption{Berry curvatures and pseudo spin textures in momentum space,
(a) for the system described by the upper block of Eq.~(\ref{eq:H_nonDirac}) with a meron of vorticity two at $\Gamma$ point,
(b) for honeycomb lattice with two merons of vorticity one at K and K$'$ points, and (c) for the band inversion scheme with a skyrmion at $\Gamma$ point.
\label{fig:modelfigs}}
\end{figure*}

\vspace{3mm}
\noindent\textit{Model Hamiltonian ---}
Let us begin with a $k\cdot p$ model at $\Gamma$ point for a system characterized by $C_{\rm 3}$
crystal symmetry in additional to time reversal symmetry, with the Bi atoms on triangle lattice in mind.
The four-band $k\cdot p$ model around $\Gamma$ point up to the second order of momentum has the following form
\begin{equation}
H(\textbf{k})=ak^2+\begin{bmatrix}\lambda_{\rm so} & \gamma k_{-}^{2} & 0 & 0\\
\gamma^* k_{+}^{2} & -\lambda_{\rm so} & 0 & 0\\
0 & 0 & -\lambda_{\rm so} & \gamma k_{-}^{2}\\
0 & 0 & \gamma^* k_{+}^{2} & \lambda_{\rm so}
\end{bmatrix},\label{eq:H_nonDirac}
\end{equation}
on basis
\{
$|p_+,\uparrow\rangle $,
$|p_-,\uparrow\rangle $,
$|p_+,\downarrow\rangle $,
$|p_-,\downarrow\rangle $
\},
with
$|p_{\pm}\rangle=\mp(|p_x\rangle\pm i |p_y\rangle)$ and
$k_{\pm}=k_x\pm ik_y$.
Time reversal symmetry requires the matrix element
between them to be an even function of momentum.  The $C_3$ rotation symmetry further
requires the matrix element between $|p_{+}\rangle$ and $|p_{-}\rangle$
to take the form of $k_{\pm}^2$ in order to conserve the total angular
moment along $z$ direction.  Therefore, the electrons in the present system are of non-Dirac type. As for the diagonal masses, linear terms in momentum are not allowed since they would
change the angular momentum of the basis wave-functions; at the second order of momentum, time reversal
symmetry does not permit any asymmetric contribution in $|p_{+}\rangle$ and $|p_{-}\rangle$ in the
individual 2$\times$2 blocks.
 The dispersion of Hamiltonian Eq.~(\ref{eq:H_nonDirac}) is then given by
$E(k)=ak^2\pm \sqrt{\lambda_{\rm so}^2+|\gamma|^2 k^4}$, which exhibits a degeneracy and quadratic band touching
at $\Gamma$ point when SOC is absent, a property originated from the property that $p_x$ and $p_y$ belong to a two-dimensional (2D) irreducible
representation to $C_3$ crystal symmetry along with time reversal symmetry.

A finite SOC removes this degeneracy and opens a gap, which may result in topologically
nontrivial states.
Since the atomic SOC $H_{\rm so}=\lambda_{\rm so} \textbf{s}\cdot\textbf{L}$
is diagonal in this basis, one can first analyze the Berry curvature for the spin-up subspace\cite{XD_NQ_RMP}
\begin{equation}
\Omega(\textbf{k})
\equiv \partial_{k_{x}}A_{y}(\textbf{k})-\partial_{k_{y}}A_{x}(\textbf{k})
=\frac{2|\gamma|^{2}\lambda_{\rm so}k^{2}}{(|\gamma|^{2}k^4+\lambda_{\rm so}^{2})^{3/2}},\label{eq:omega}
\end{equation}
where the Berry connection is defined as
$A_{\alpha}(\mathbf{k})
=i\langle u(\mathbf{k})|\partial_{k_{\alpha}}|u(\mathbf{k})\rangle$ and
$|u(\textbf{k})\rangle$ is the occupied state. The dispersion of
$\Omega(\textbf{k})$ in momentum space is plotted in Fig.~\ref{fig:modelfigs}(a).
The Chern number is evaluated as\cite{TKNN_1982,XD_NQ_RMP}
\begin{equation}
C_{\uparrow}=\frac{1}{2\pi}\int d\mathbf{k}\Omega(\mathbf{k})
={\rm sign}(\lambda_{\rm so}).
\label{eq:ChernNumber}
\end{equation}
This Chern number can be understood from the topological texture of pseudo spin
referring to the two orbits in the upper block of Eq.~(\ref{eq:H_nonDirac})
$
\bm{\sigma}
=(-1)\{\gamma(k_{x}^{2}-k_{y}^{2}),\;2\gamma k_{x}k_{y},\;\lambda_{\rm so}\}/\sqrt{\gamma^{2}k^{4}+\lambda_{\rm so}^{2}},
$
as shown in Fig.~\ref{fig:modelfigs}(a)
(a gauge is taken to make $\gamma$ real).
At $\Gamma$ point, the pseudo spin is along the north pole; as $k$ increases,
the pseudo spin vector tilts away and becomes aligned to the equator (namely the $k_x-k_y$ plane).
Checking the rotation of pseudo spin, we find a
$4\pi$ rotation around $\Gamma$ point. This meron texture with vorticity
two contributes the unity Chern number in Eq.~(\ref{eq:ChernNumber}).
With time reversal symmetry, one can see the same physics in the
spin-down channel. The topology of the whole system is therefore described by
the $Z_2$ index\cite{Kane_Z2_2005,FuLiang_3DTI_2007}, which {\color{purple} is} defined by the spin Chern numbers as
$
Z_{2}=\frac{1}{2}(C_{\uparrow}-C_{\downarrow})\; mod\;2.
$
Because the particle-hole symmetry is broken by the quadratic term $a k^2$ in Eq.~(\ref{eq:H_nonDirac}),
the present system falls into the class AII\cite{Kitaev_2009,Ryu_2010}.

It is noticed that SOC contributes totally in opening the topological gap in the present scheme,
since $C_\uparrow$ (and thus $Z_2$ index) is always one for nonzero SOC as seen in Eq.~(\ref{eq:ChernNumber}).  This is similar to the case of honeycomb lattice in
the Haldane model and Kane-Mele model \citep{Haldane_1988,Kane_Mele_Graphene_PRL_2005}, where tuning the signs of
mass terms at K and K$^\prime$ adequately aligns the pseudo spin texture in the two merons (see Fig.~\ref{fig:modelfigs}(b)), and results in various topological states
with nonzero integer topological charge in the whole Brillouin zone (BZ) \cite{LWH_2013}. In contrary, the band inversion in BHZ model \citep{Bernevig_HgTe_QW_science_2006}
is associated with a skyrmion of the pseudo spin texture (see Fig.~\ref{fig:modelfigs}(c)), where the topological gap is determined by the hybridization of two inverted bands.

\begin{figure}[t]
\centering{}
\includegraphics[width=0.48\textwidth]{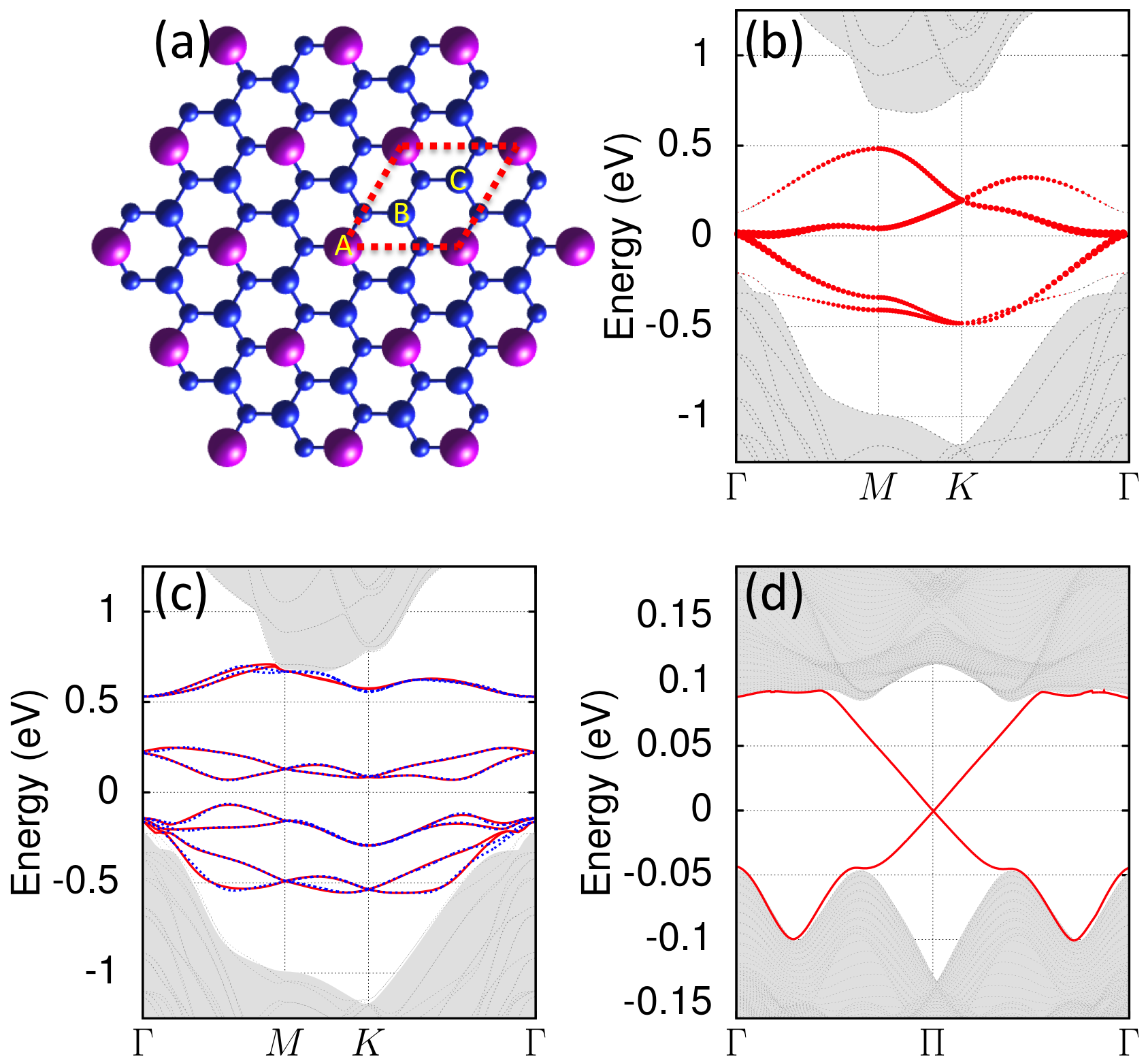}
\caption{(Color online.)
(a) Top view of the crystal structure for 1/3 Bi coverage on the Si[111] surface, with
the large and small blue balls for Si atoms of the top-most and the second layers, respectively, and purple balls for Bi atoms;
(b) and (c) electronic band structures by first principles calculations without and with SOC, respectively, where the magnitudes of Bi
$p_x$ and $p_y$ orbits are represented by the width of red curves,  and the band structures from a tight-binding Hamiltonian constructed
with maximally-localized Wannier functions (MLWF) method are shown by dashed blue curves;
(d) 1D energy bands for a ribbon system with the edge state on one of the two edges calculated based on the Wannierized Hamiltonian.
\label{fig:bivvSi} }
\end{figure}

\vspace{3mm}
\noindent\textit{Bi on Si{[}111{]} surface ---}
To be specific, we investigate the Si{[}111{]} surface with 1/3 coverage by Bi atoms. Based on symmetry analysis
and first-principles calculations, we show that this system realizes the
physics described by the model Hamiltonian Eq.~(\ref{eq:H_nonDirac}).

The structure is  schematically shown in Fig.~\ref{fig:bivvSi}(a).
On the Si{[}111{]} surface, the top-most  layer Si atoms form a triangle
lattice and the unsaturated dangling bonds of Si provide active adsorption
sites for various metal atoms.  When a $\sqrt{3}\times\sqrt{3}$
unit cell is considered, three sites (A, B and C) of Si surface can be
saturated by ad-atoms, which we call A-B-C/Si[111]. A, B and C can either be occupied by atoms such as
Bi, Pb, Au, Ag and H, or left vacant where the dangling bonds of Si atoms are unsaturated. Here
we consider explicitly the case that A sites are occupied by Bi atoms
while B and C sites remain vacant, as shown in the top view in Fig.~\ref{fig:bivvSi}(a). Bi atoms form a triangle sub-lattice and
the symmetry group is $C_{\rm 3v}$  when the Si substrate is taken into account.

First we study the electronic band structures of this system by first-principles
calculations. The calculations were done with the Vienna \emph{ab initio}
simulation package (VASP)\cite{VASP}, where projected augmented-wave (PAW)
potential is adopted\cite{PAW_Blochl,PAW_Kresse_1999}. We use the functional introduced by Perdew, Burke,
and Ernzerhof (PBE) \cite{PBE} and the generalized gradient approximation (GGA).
A 6$\times$6$\times$1 mesh in the irreducible Brillouin Zone was used for
structure relaxation and 8$\times$8$\times$1 mesh for self-consisted calculations.
In all the calculations the energy cutoff is set as 500 eV. The forces are
relaxed lower than 0.01 eV/$\rm \AA$. After relaxation, the bond length between
the surface Si and the attached Bi becomes 2.58$\rm \AA$, close to the summation
of their ion radius, indicating a strong binding interaction between two atoms. The
Bi-Bi distance is 3.84$\rm \AA$, which is also consistent with the structure of Si[111] surface. The unsaturated Si atoms,
on the other hand, form a honeycomb lattice with a Bi atom sitting at the center of each hexagon.
All the calculations are done in a slab structure with 12 layers of Si atoms, and a vacuum layer of 10 $\AA$ is included.
In order to avoid annoying physics, Si atoms on the bottom surface are saturated by H atoms.

The band structure without SOC is shown in Fig.~\ref{fig:bivvSi}(b).
The shadowed (gray color) areas are bulk bands of Si, where the indirect gap of Si crystal
is clearly seen. Inside the gap four bands are found from Bi atoms and unsaturated
surface Si atoms. The magnitude of the projection of each Bloch states onto the
Bi $p_x$ and $p_y$ orbits is represented by the width of red curves.
A quadratical touching is found between the second and third bands at $\Gamma$ point. The Fermi
energy passes through this degenerate point and makes the system a
zero gap semiconductor. A gap is opened when SOC is turned on as shown in Fig.~\ref{fig:bivvSi}(c).
The degeneracy at $\Gamma$ point originally of four-fold due to the
orbital and spin degrees of freedom is lifted partially down to two-fold.
Away from the high symmetry point the Rashba SOC due to the broken inversion symmetry at surface
lifts the degeneracy totally, which also shifts the position of minimal gap
to a midpoint in between $\Gamma$ and $M$ points.

In order to understand the possible topological property of the system, we
construct the effective model Hamiltonian around $\Gamma$ point
\begin{equation}
H_{\Gamma}=\left[\begin{array}{cccc}
a_{1}k^{2} & 0 & c_{2}k_{-} & -c_{2}^{*}k_{-}\\
 & a_{1}k^{2} & c_{2}^{*}k_{+} & -c_{2}k_{+}\\
 &  & \epsilon_{z}+a_{2}k^{2} & d_{0}\\
\dagger &  &  & \epsilon_{z}+a_{2}k^{2}
\end{array}\right],
\label{eq:H_Gamma_pxpypzpz}
\end{equation}
on the basis of four orbits
$\{|p_+\rangle$,
$|p_-\rangle$,
$|p_z^B\rangle$,
$|p_z^\textsc{C}\rangle
\}$
inside the gap of bulk Si system, taking into account $C_{\rm 3v}$ crystal symmetry and time reversal symmetry.
The parameters are evaluated by fitting the results of first-principles calculations as
$\epsilon_z=0.01$eV, $a_1=0.55$eV$\cdot\rm \AA^{2}$, $a_2=-0.6$eV$\cdot\rm \AA^{2}$,  $c_2=0.3i$eV$\cdot\rm \AA$ and $d_0=0.18$eV.

Around $\Gamma$ point, the fat-band structure in Fig.~\ref{fig:bivvSi}(b) shows that the two states close to Fermi energy are
mainly from Bi $|p_{\pm}\rangle$ orbits and the other two states
far away from Fermi energy are contributed mainly by the
two $p_z$ orbits of the unsaturated Si atoms. We therefor can down-fold this 4$\times$4 Hamiltonian into the
subspace spanned by $|p_{\pm}\rangle$ orbits. Taking the spin degree of freedom and SOC
into consideration, we obtain the 4$\times$4 effective Hamiltonian as shown in
Eq.~(\ref{eq:H_nonDirac}). Therefore, the system is a QSHE insulator as discussed above.

An important character of the QSHE insulating state is the existence of gapless states at the sample edge. We calculate the band structure
for a ribbon of width of 50 unit cells based on the maximally localized Wannier function for the surface electronic states\cite{Vanderbilt_RMP},
where Rashba SOC is included. As shown in Fig.~\ref{fig:bivvSi}(d), there are
two gapless edge states connecting the valence and conduction bands for one of the two edges of the ribbon, which cross
at the time-reversal invariant point in the 1D BZ. This indicates that this system
is in a topologically nontrivial phase with a topological index  $Z_2=1$.
It is noticed that the QHSE state survives even
in the presence of Rashba SOC, although the global gap is reduced partially.

\begin{figure}[t]
\centering{}
\includegraphics[width=0.47\textwidth]{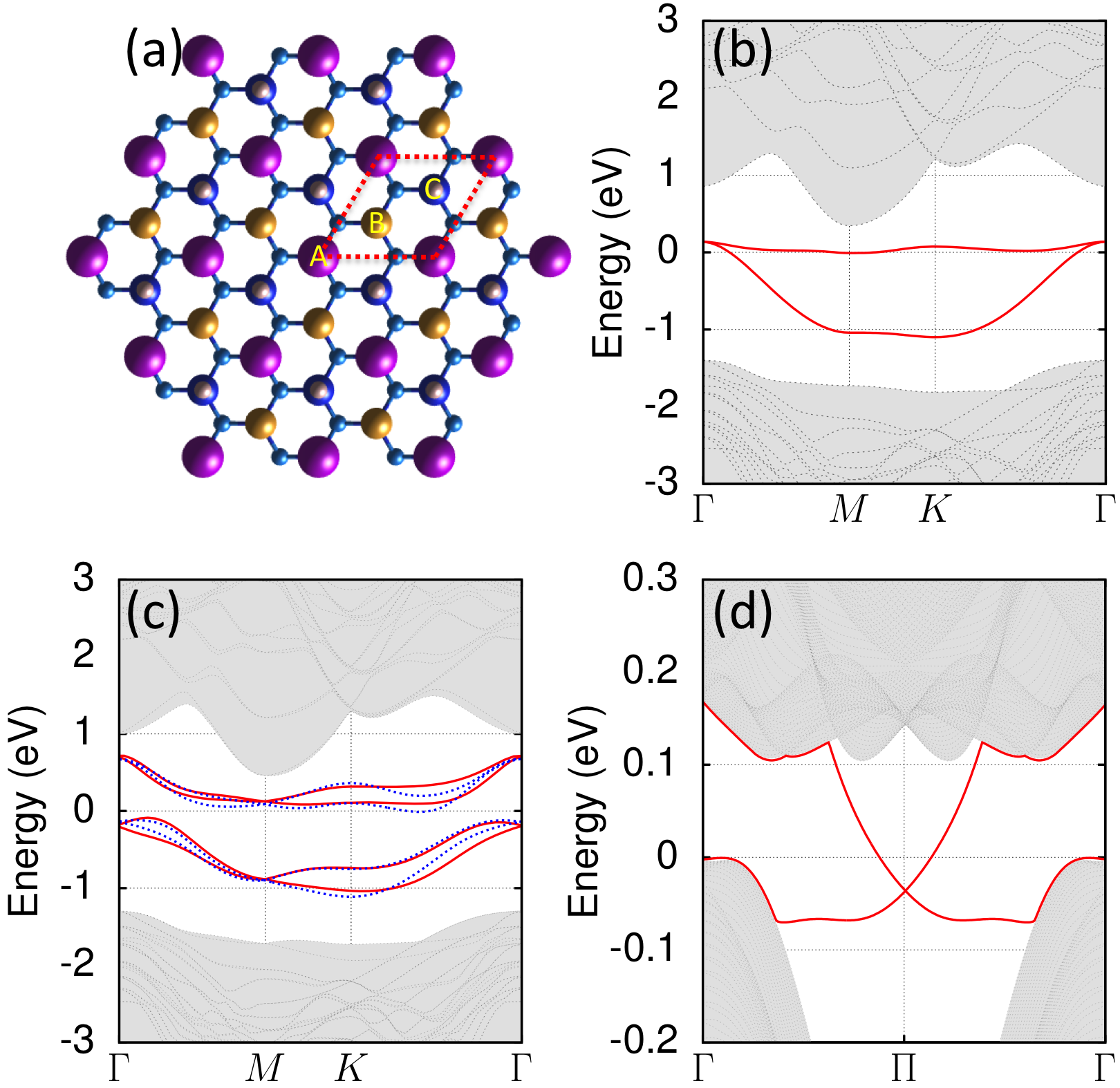}
\caption{(Color online.)
 (a) Top view of the crystal structure for Bi-Au-H/SiC{[}111{]} surface where the small blue balls for Si atoms of the second layer, purple balls for Bi atoms,
 gold balls for Au atoms, silver balls for H atoms, respectively;
  (b) and (c) electronic band structures by first principles calculations, without and with SOC respectively; (d) 1D energy bands for a ribbon system.
\label{fig:BiAuHSiC}}
\end{figure}

\vspace{3mm}
\noindent\textit{Bi-Au-H on SiC{[}111{]} surface ---}
As a second example, we consider the system Bi-Au-H/SiC{[}111{]} where a SiC substrate with smaller lattice
constant of 3.06$\AA$ is used as compared with 3.84$\AA$ for the Si substrate.  As can be read from Fig.~\ref{fig:BiAuHSiC}(a), the crystal symmetry is lowered to $C_3$.
Since the $p_z$ orbit of Si atoms is now
saturated by Au and H atoms, there are only $p_x$ and $p_y$ orbits in the bulk gap as seen in Fig.~\ref{fig:BiAuHSiC}(b).
The degeneracy at $K$ point is lifted since the mirror symmetry is absent, while that at $\Gamma$ point is protected (in absence of SOC) by
time reversal symmetry due to the mechanism known for long time\cite{Herring_1937}. Therefore, the $k\cdot p$ model Eq.~(\ref{eq:H_nonDirac}) applies
to the present system, and a QHSE is realized as evidenced by the gap opening at $\Gamma$ point upon introducing SOC (Fig.~\ref{fig:BiAuHSiC}(c)) and the gapless edge state in
a ribbon structure (Fig.~\ref{fig:BiAuHSiC}(d)).

\vspace{3mm}
\noindent\textit{Discussions and summary---}
Since $d_{\pm}=d_{xz}\pm id_{yz}$ orbits (or corresponding $f$ orbits) are transformed to each other in the same manner as the $p_{\pm}$ orbits under the
$C_3$ (and/or $C_6$) rotation, QSHE states are also possible with the $d_{\pm}$ subbands,
supposing other $d$ orbits (or $f$ orbits) can be filtered in some way \cite{XuYong_ZhangSC_BiH_PRL_2013,
LiuFeng_BiH_Si_2014,Yao_BiX_SbX_2Dhoneycomb}.

Let us inspect the possibility of topologically nontrivial states featured by meron with vorticity higher than two.
In order to generate a meron with vorticity $2n$, the lowest order of momentum in the off-diagonal elements in Eq.~(\ref{eq:H_nonDirac}) should be $k_{\pm}^{2n}$, which links the two basis wave-functions with the angular-momentum difference $\pm 2n$. Meanwhile, the basis wave-functions should belong to a 2D irreducible
representation of the space groups of crystal,
which protects the degeneracy at $\Gamma$ point in absence of SOC.
Let us see the case $d_{\pm2}=d_{x^2-y^2}\pm 2id_{xy}$, with the difference in angular momentum between the basis wave-functions equal to four. In a crystal with $C_3$ (or $C_6$) symmetry, the lowest order of momentum in the off-diagonal terms preserving the total angular momentum are $k_{\pm}^2$, rather than $k_{\pm}^4$, which generates a meron with vorticity two
as discussed above. On the other hand, for a crystal with $C_4$ symmetry, the two wave functions do not form a 2D irreducible representation.
Therefore, the meron texture with vorticity two addressed in the present work is unique in generating topological states.

Finally we discuss briefly non-Dirac dispersions in other topological states.
In a Chern semimetal HgCr$_2$Se$_4$, the off-diagonal interaction has the
form of $k_zk^2_{\pm}$ in the effective $k\cdot p$ Hamiltonian \cite{XuGang_Dai_HgCrSe}. In their case, there is a quadratic momentum dependence
in the diagonal mass terms, which makes the pseudo spin texture not a meron.
Quadratically touching band structures at $\Gamma$ point appear in multi-band systems on honeycomb lattice
as discussed in a model system recently, due to the same $C_3$ symmetry \cite{CJW_2014}. Because the non-Dirac dispersion is deep in
the valence band, it does not contribute directly to the nontrivial topology.

To summarize, we propose the Si{[}111{]} surface with Bi atoms adapted as a new template for achieving topological states.
With the $C_3$ crystal symmetry and time reversal symmetry, a $k\cdot p$ model is constructed with degeneracy at $\Gamma$ point
and quadratic, non-Dirac type energy dispersion. When the spin-orbit coupling is taken into account,
a gap is opened resulting in a quantum spin Hall effect insulator as characterized by a meron pseudo spin texture with vorticity
two, in contrast to the widely studied Dirac electrons. In the present scheme, the strong atomic spin-orbit coupling of Bi contributes fully to
open the topological gap, and thus provides a new possibility for achieving topological states with large gap which may function even above
room temperature. With the Si substrate taking part in realizing the topological state, the present template is expected to
make the integration of topological states into existing electronics and photonics devices very promising.

\vspace{3mm}
\noindent\textit{Acknowledgements ---} The authors are grateful to T.~Uchihashi and T.~Kawakami for stimulating discussions.
This work was supported by WPI Initiative on Materials Nanoarchitectonics, MEXT, Japan, and
the Grant-in-Aid for Scientific Research under the Innovative Area ``Topological Quantum Phenomen'' (No.25103723), MEXT, Japan.
Q.F.L acknowledges the support from NSFC under grants 10904092.

Q.F.L. and Y.R. contributed equally to this work.

\bibliographystyle{apsrev4-1}
\bibliography{refs}

\end{document}